\documentclass[12pt]{article}
\pdfoutput=1
\usepackage{graphicx}
\usepackage{amsmath}
\usepackage{amssymb}
\usepackage{cite}
\def\slashchar#1{\setbox0=\hbox{$#1$} 
\dimen0=\wd0 
\setbox1=\hbox{/} \dimen1=\wd1 
\ifdim\dimen0>\dimen1 
\rlap{\hbox to \dimen0{\hfil/\hfil}} 
#1 
\else 
\rlap{\hbox to \dimen1{\hfil$#1$\hfil}} 
/ 
\fi}

\def\kahler{K\"ahler\ }
\begin{document}
\def\a{\alpha}
\def\b{\beta}
\def\c{\varepsilon}
\def\d{\delta}
\def\e{\epsilon}
\def\f{\phi}
\def\g{\gamma}
\def\h{\theta}
\def\k{\kappa}
\def\l{\lambda}
\def\m{\mu}
\def\n{\nu}
\def\p{\psi}
\def\q{\partial}
\def\r{\rho}
\def\s{\sigma}
\def\t{\tau}
\def\u{\upsilon}
\def\v{\varphi}
\def\w{\omega}
\def\x{\xi}
\def\y{\eta}
\def\z{\zeta}
\def\D{{\mit \Delta}}
\def\G{\Gamma}
\def\H{\Theta}
\def\L{\Lambda}
\def\F{\Phi}
\def\P{\Psi}

\def\S{\Sigma}

\def\o{\over}
\def\beq{\begin{eqnarray}}
\def\eeq{\end{eqnarray}}
\newcommand{\gsim}{ \mathop{}_{\textstyle \sim}^{\textstyle >} }
\newcommand{\lsim}{ \mathop{}_{\textstyle \sim}^{\textstyle <} }
\newcommand{\vev}[1]{ \left\langle {#1} \right\rangle }
\newcommand{\bra}[1]{ \langle {#1} | }
\newcommand{\ket}[1]{ | {#1} \rangle }
\newcommand{\EV}{ {\rm eV} }
\newcommand{\KEV}{ {\rm keV} }
\newcommand{\MEV}{ {\rm MeV} }
\newcommand{\GEV}{ {\rm GeV} }
\newcommand{\TEV}{ {\rm TeV} }
\def\diag{\mathop{\rm diag}\nolimits}
\def\Spin{\mathop{\rm Spin}}
\def\SO{\mathop{\rm SO}}
\def\O{\mathop{\rm O}}
\def\SU{\mathop{\rm SU}}
\def\U{\mathop{\rm U}}
\def\Sp{\mathop{\rm Sp}}
\def\SL{\mathop{\rm SL}}
\def\tr{\mathop{\rm tr}}

\def\IJMP{Int.~J.~Mod.~Phys. }
\def\MPL{Mod.~Phys.~Lett. }
\def\NP{Nucl.~Phys. }
\def\PL{Phys.~Lett. }
\def\PR{Phys.~Rev. }
\def\PRL{Phys.~Rev.~Lett. }
\def\PTP{Prog.~Theor.~Phys. }
\def\ZP{Z.~Phys. }

\newcommand{\drawsquare}[2]{\hbox{%
\rule{#2pt}{#1pt}\hskip-#2pt
\rule{#1pt}{#2pt}\hskip-#1pt
\rule[#1pt]{#1pt}{#2pt}}\rule[#1pt]{#2pt}{#2pt}\hskip-#2pt
\rule{#2pt}{#1pt}}

\def\vbr{\vphantom{\sqrt{F_e^i}}}
\newcommand{\fund}{\drawsquare{6.5}{0.4}}
\newcommand{\afund}{\overline{\fund}}
\newcommand{\symm}{\drawsquare{6.5}{0.4}\hskip-0.4pt%
        \drawsquare{6.5}{0.4}}
\newcommand{\asymm}{\raisebox{-3pt}{\drawsquare{6.5}{0.4}\hskip-6.9pt%
        \raisebox{6.5pt}{\drawsquare{6.5}{0.4}}}}
\newcommand{\asymmthree}{\raisebox{-7pt}{\drawsquare{6.5}{0.4}}\hskip-6.9pt%
\raisebox{-0.5pt}{\drawsquare{6.5}{0.4}}\hskip-6.9pt%
\raisebox{6pt}{\drawsquare{6.5}{0.4}}}
\newcommand{\asymmfour}{\raisebox{-10pt}{\drawsquare{6.5}{0.4}}\hskip-6.9pt%
\raisebox{-3.5pt}{\drawsquare{6.5}{0.4}}\hskip-6.9pt%
\raisebox{3pt}{\drawsquare{6.5}{0.4}}\hskip-6.9pt%
        \raisebox{9.5pt}{\drawsquare{6.5}{0.4}}}
\newcommand{\Ythrees}{\raisebox{-.5pt}{\drawsquare{6.5}{0.4}}\hskip-0.4pt%
          \raisebox{-.5pt}{\drawsquare{6.5}{0.4}}\hskip-0.4pt%
          \raisebox{-.5pt}{\drawsquare{6.5}{0.4}}}
\newcommand{\Yfours}{\raisebox{-.5pt}{\drawsquare{6.5}{0.4}}\hskip-0.4pt%
          \raisebox{-.5pt}{\drawsquare{6.5}{0.4}}\hskip-0.4pt%
          \raisebox{-.5pt}{\drawsquare{6.5}{0.4}}\hskip-0.4pt%
          \raisebox{-.5pt}{\drawsquare{6.5}{0.4}}}
\newcommand{\Ythreea}{\raisebox{-3.5pt}{\drawsquare{6.5}{0.4}}\hskip-6.9pt%
        \raisebox{3pt}{\drawsquare{6.5}{0.4}}\hskip-6.9pt
        \raisebox{9.5pt}{\drawsquare{6.5}{0.4}}}
\newcommand{\Yfoura}{\raisebox{-3.5pt}{\drawsquare{6.5}{0.4}}\hskip-6.9pt%
        \raisebox{3pt}{\drawsquare{6.5}{0.4}}\hskip-6.9pt
        \raisebox{9.5pt}{\drawsquare{6.5}{0.4}}\hskip-6.9pt
        \raisebox{16pt}{\drawsquare{6.5}{0.4}}}
\newcommand{\Yadjoint}{\raisebox{-3.5pt}{\drawsquare{6.5}{0.4}}\hskip-6.9pt%
        \raisebox{3pt}{\drawsquare{6.5}{0.4}}\hskip-0.4pt
        \raisebox{3pt}{\drawsquare{6.5}{0.4}}}
\newcommand{\Ysquare}{\raisebox{-3.5pt}{\drawsquare{6.5}{0.4}}\hskip-0.4pt%
        \raisebox{-3.5pt}{\drawsquare{6.5}{0.4}}\hskip-13.4pt%
        \raisebox{3pt}{\drawsquare{6.5}{0.4}}\hskip-0.4pt%
        \raisebox{3pt}{\drawsquare{6.5}{0.4}}}
\newcommand{\Yflavor}{\Yfund + \overline{\Yfund}} 
\newcommand{\Yoneoone}{\raisebox{-3.5pt}{\drawsquare{6.5}{0.4}}\hskip-6.9pt%
        \raisebox{3pt}{\drawsquare{6.5}{0.4}}\hskip-6.9pt%
        \raisebox{9.5pt}{\drawsquare{6.5}{0.4}}\hskip-0.4pt%
        \raisebox{9.5pt}{\drawsquare{6.5}{0.4}}}%

\baselineskip 0.7cm

\begin{titlepage}
\begin{figure}[t]
 \begin{flushright}
  IPMU10-0140
 \end{flushright}
\end{figure}

\vskip 1.35cm
\begin{center}
{\large \bf
Cascade supersymmetry breaking\\ and\\ low-scale gauge mediation
}
\vskip 1.2cm
Masahiro Ibe${}^{1}$, Yuri Shirman${}^{1}$ and Tsutomu T. Yanagida${}^{2}$

\vskip 0.4cm
${}^{1}$ 
{\it Department of Physics \& Astronomy, University of California, Irvine, CA 92697, USA}\\
${}^2$
{\it Institute for the Physics and Mathematics of the Universe, University of Tokyo, \\
Kashiwa 277-8568, Japan}

\vskip 1.5cm

\abstract{
We propose a new class of models with gauge mediated supersymmetry breaking, the cascade supersymmetry breaking.
This class of models is consistent with the gravitino mass as low as $O(1)$\,eV without having suppressed gaugino masses,
nor the Landau pole problems of the gauge coupling constants of the Standard Model
below the scale of the grand unification.
In particular, there is no supersymmetric vacuum in the vicinity of the supersymmetry breaking vacuum 
even for such a low gravitino mass.
Thus, the model does not have a vacuum stability problem decaying into supersymmetric vacua.
}
\end{center}
\end{titlepage}

\setcounter{page}{2}

\section{Introduction}
The gauge mediated supersymmetry breaking (GMSB) models\,\cite{Dine:1981za,Dine:1981gu,Dimopoulos:1982gm,
Affleck:1984xz,Dine:1993yw,Dine:1994vc,Dine:1995ag} provide one of the most 
attractive realizations of the phenomenologically
acceptable minimal supersymmetric standard model (MSSM).
The gravitino is the  lightest supersymmetric particle whose 
mass ranges from the eV to the GeV.
This feature of gauge mediation motivated a lot of theoretical works on
phenomenological and cosmological aspects of the models\,\cite{Feng:2010ij}.

Light gravitino models are constrained by cosmological and astrophysical problems. In particular, the analysis of  CMB data and the Lyman-$\alpha$ forest data put an upper bound of 16\,eV on the gravitino mass\,\cite{Viel:2005qj}. Furthermore, future cosmic microwave background observations will probe gravitino mass down to the eV range\,\cite{Ichikawa:2009ir}. 
Thus, models with very light gravitino are particularly interesting. Such models require a very low fundamental scale of supersymmetry (SUSY) breaking and as a result may have very rich collider phenomenology.

Models with such a light gravitino (see for example 
Ref.\cite{Izawa:1997gs,Csaki:2006wi,Dine:2006xt,Shirai:2010rr,Kitano:2010fa}) generically suffer from a range of theoretical problems: the Standard Model couplings hit the Landau pole below the scale of the Grand Unification Theory (GUT); 
gaugino masses are suppressed compared to the sfermion masses; there are supersymmetric ground states in the vicinity of desired SUSY breaking vacua.
Much of the parameter space in this class of models has been excluded by the 
neutralino/chargino mass limit\,\cite{Ibe:2005xc,Sato:2009dk,Sato:2010tz} 
at the Tevatron experiments\,\cite{Aaltonen:2009tp} 
or by the vacuum stability problem\,\cite{Hisano:2008sy} for the gravitino lighter than $16$\,eV.
In this paper, we propose a new class of the GMSB models,  
the cascade supersymmetry breaking, which may avoid these problems.

The organization of the paper is as follows.
In section\,\ref{sec:GMSB}, we briefly review
problems arising in attempts to construct low scale GMSB models.
In section\,\ref{sec:cascade},  we introduce a generic idea of the cascade supersymmetry breaking.
An example based on dynamical supersymmetry breaking (DSB) is introduced in section\,\,\ref{sec:model}.
In sections\,\,\ref{sec:spectrum}, we study phenomenological features of the model, messenger and superpartner spectra as well as possible  generalizations.
In section\,\ref{sec:conformal}, we comment 
on a relation of our model to
the conformal gauge mediation mechanism developed in Ref.\cite{Ibe:2007wp}.
The final section is devoted to the conclusions and some discussions.

\section{A brief review of low scale gauge mediation}\label{sec:GMSB}

\subsection{Scales in light gravitino scenario}
Since we are interested in models with the gravitino mass in the eV range, the fundamental SUSY breaking scale must be low,
\begin{eqnarray}
\label{eq:Fgravitino}
\sqrt{F} \sim 65\,{\rm TeV} \times \left( \frac{m_{3/2}}{1\,{\rm eV}}\right)^{1/2}\ .
\end{eqnarray}
Such a low SUSY breaking scale can be achieved in GMSB models. Indeed, in the simplest GMSB models, the superpartner masses are given by
\begin{eqnarray}
 \tilde m \sim \frac{\alpha}{4\pi} \frac{F_S}{m}\,,
\end{eqnarray}
where $m$ is the messenger scale and $F_S$ is a mass splitting within messenger multiplets.
One must also require $F_S<m^2$ to avoid tachyonic messengers and charge-color breaking. 
Thus, a requirement that the superpartners have mass at the electroweak scale leads to the lower bound on the mass parameters in the messenger sector,
\begin{eqnarray}
 m \sim \sqrt{F_S} \sim O(10-100)\,{\rm TeV}\ .
\end{eqnarray}
We see that light gravitino scenarios can only be realized when the SUSY breaking effects in the messenger sector are comparable to the fundamental scale of SUSY breaking, $F_S\sim F$. To avoid the separation of scales, one must look at models of direct or semi-direct gauge mediation. Furthermore, one expects that successful models will necessarily be strongly coupled.

\subsection{R-symmetry and the messenger sector}
Let us briefly review difficulties encountered in search for models of low scale gauge mediation.
We begin by considering a simple example of the messenger sector\footnote{Low energy description of direct GMSB models is often given by superpotentials of this type \cite{Izawa:1997gs,Csaki:2006wi}.}:
\begin{eqnarray}
\label{eq:rsymmessenger}
 W  = \Lambda_{1}^2 S + (m_{ij} + \l_{ij} S ) \bar{\psi}_i \psi_j \ ,
\end{eqnarray}
Where $S$ is a supersymmetry breaking field and $\psi$, $\bar\psi$ are messenger fields.
This model possesses an R-symmetry (under which $S$ has charge 2) if the charges of the messenger fields can be chosen such that\,\cite{Shih:2007av,Komargodski:2009jf}:
\begin{eqnarray}
\label{eq:rsymconditions}
m_{ij}\ne 0 ~\Rightarrow ~R(\psi_i)+R(\psi_j)=2;\hskip 0.5in \lambda_{ij}\ne 0~\Rightarrow~R(\psi_i)+R(\psi_j)=0\,.
\end{eqnarray}
We will restrict our attention to models with $\det m \ne 0$. 
Indeed, models with $\det m = 0$ are problematic from phenomenological 
perspective because  the true vacuum in this case is a charge-color breaking 
one (for detailed analysis of the vacuum structure of this class of models see 
Ref.\cite{Komargodski:2009jf}). When parameters are chosen so that the gravitino mass is in the 
$\mathcal{O}(10\,\mathrm{eV})$ range, the charge-color breaking ground states 
are found in the vicinity of the charge-color preserving minima of the 
potential. This leads to two problems with the desired minimum: first, the 
lifetime of this minimum is expected to be too short;  second, initial 
conditions must be fine-tuned for it to be selected in the course of 
cosmological evolution.%
\footnote{
A very simple model can be constructed\,\cite{Ibe:2007ab} 
if the requirement of a very light gravitino is relaxed.
See also Ref.\cite{Giveon:2009yu}
for models with $\det m = 0$.}

Furthermore, the lifetime of the vacuum in the models with light gravitino 
and supersymmetric runaway directions is too short. We will, therefore, 
require that $\lambda m^{-1}\lambda=0$ which guarantees the absence of the runaway
directions\,\cite{Sato:2009dk}.
With these assumptions, the model in Eq.(\ref{eq:rsymmessenger}) possesses
the following important features
\begin{itemize}
 \item The supersymmetry is broken;%
 \item The effective mass matrix for messenger fields is independent of the 
 vacuum expectation value (vev)  of  the pseudo-modulus: 
\begin{eqnarray}
\label{eq:Sindependent}
\det (m_{ij}  + \l_{ij} S ) = \det m_{ij}\neq 0\ .
\end{eqnarray}
\end{itemize}

A phenomenologically viable model requires that the R-symmetry is broken (otherwise the Standard Model gauginos remain massless). This can be achieved through introduction of additional fields and interactions that lead to either spontaneous or explicit R-symmetry breaking%
\footnote{Such interactions generically lead to appearance of supersymmetric vacua elsewhere on the moduli space. We will discuss problems associated with the existence of such vacua shortly.}. 
However, this is often insufficient. Indeed, as long as the effective messenger mass matrix satisfies 
Eq.(\ref{eq:Sindependent}), 
the gaugino masses vanish at the leading order in $F_S$\,\cite{Giudice:1997ni}
\begin{eqnarray}
\label{eq:leading}
m_{\tilde g} =\left. 
\frac{\alpha }
{4\pi} \log\left[\det (m_{ij}  + \l_{ij} S )\right]\right|_{\rm \theta^2}=0\ .
\end{eqnarray}
The first contribution to gaugino masses appears only at the order $F_S^3$\,\cite{Izawa:1997gs},
\begin{eqnarray}
 m_{\rm gaugino} \sim c \frac{\a}{4\pi} \frac{F_S}{m}\left|\frac{F_S}{m^2}\right|^2\ .
\end{eqnarray}
The sfermion masses are still generated at the leading order in $F_S$ and the
requirement that gaugino and sfermion masses are comparable can only be satisfied in models of low scale SUSY breaking. However, it turns out that the coefficient $c$ is small and sufficiently large gaugino masses can not be achieved without fine-tuning.
Furthermore, detailed numerical analysis, has shown that, 
even with $c= O(1)$, the predicted gaugino masses  have been almost excluded by Tevatron
constraints on the neutralino/chargino masses
for $m_{3/2}\lesssim16$\,eV\,\cite{Ibe:2005xc, Sato:2009dk,Sato:2010tz}.

Thus, we must turn to models  where Eq.(\ref{eq:rsymconditions}) is not satisfied in the messenger sector.
In such models, gaugino masses are unsuppressed, however, the SUSY breaking vacuum is only a local one.
To see this, note that at least at one point along the pseudo-flat direction the matrix $m_{ij}+\lambda_{ij}S$ has a zero eigenvalue and at least one pair of messengers becomes massless. They can now acquire vevs and restore supersymmetry.
To ensure sufficiently long vacuum lifetime, one must increase the messenger mass $m$. 
This suppresses all superpartner masses if the gravitino mass (and, therefore,  the fundamental scale of SUSY breaking) is kept fixed.
In particular, for $m_{3/2}\lesssim 16$\,eV, the detailed numerical analysis \,\cite{Hisano:2008sy} has lead to an upper bound on superpartner masses of about $1$\,TeV.

\subsection{Direct and semi-direct gauge mediation}
A toy model discussed above implied direct or semi-direct gauge mediation. Additional problems often arise when UV complete realizations of direct gauge mediation is considered (a precise definition of direct gauge mediation is given in \cite{Dine:2007dz,Carpenter:2008wi} while semi-direct gauge mediation is introduced in \cite{Seiberg:2008qj}).
The messengers play a dual role in direct GMSB models --- in addition to communicating the supersymmetry breaking to the Standard Model fields, they play a role in SUSY breaking dynamics.
The need for a large flavor symmetry in the DSB sector usually implies that the DSB gauge group is large itself.
As a result, the Standard Model gauge interactions have 
Landau poles below the GUT scale when the messenger scale is low%
\footnote{The Landau pole problems may be ameliorated 
if the messenger fields receive large positive anomalous
dimensions under the renormalization group evolution 
from the GUT to the messenger scale\,\cite{Sato:2009yt}.
}.
If one insists on perturbative coupling unification, both the messenger and SUSY breaking scales are pushed up preventing the possibility of light gravitino.
This latter difficulty may be avoided in semi-direct gauge mediation\,\cite{Ibe:2007wp,Seiberg:2008qj} 
(see also Refs.\cite{Randall:1996zi,Izawa:2005yf} for  earlier attempts), where messengers are charged under the DSB group but do not play a direct role in SUSY breaking. This allows to construct models with small number of messengers and avoid Landau pole problems.
However, the leading contribution to the gaugino mass is again
vanishing due to the R-symmetry (see also Ref.\cite{Izawa:2008ef}).

 \section{Cascade supersymmetry breaking}\label{sec:cascade}
In order to solve problems discussed in the previous section we will employ some of the tools proposed in the original GMSB models\cite{Dine:1993yw,Dine:1994vc,Dine:1995ag} -- namely we will introduce a secondary SUSY breaking sector. Despite the existence of several sectors, our model will have all the desirable features of direct (and semi-direct) gauge mediation. Finally, in a strong coupling limit, we will be able to obtain low SUSY breaking scale. We will refer to this class of models as cascade SUSY breaking\,\footnote{
The cascade supersymmetry is also implemented in 
Refs.\cite{Nomura:1997ur,Fujii:2003iw}.
}.

\begin{figure}[t]
\begin{center}
  \includegraphics[width=.8\linewidth]{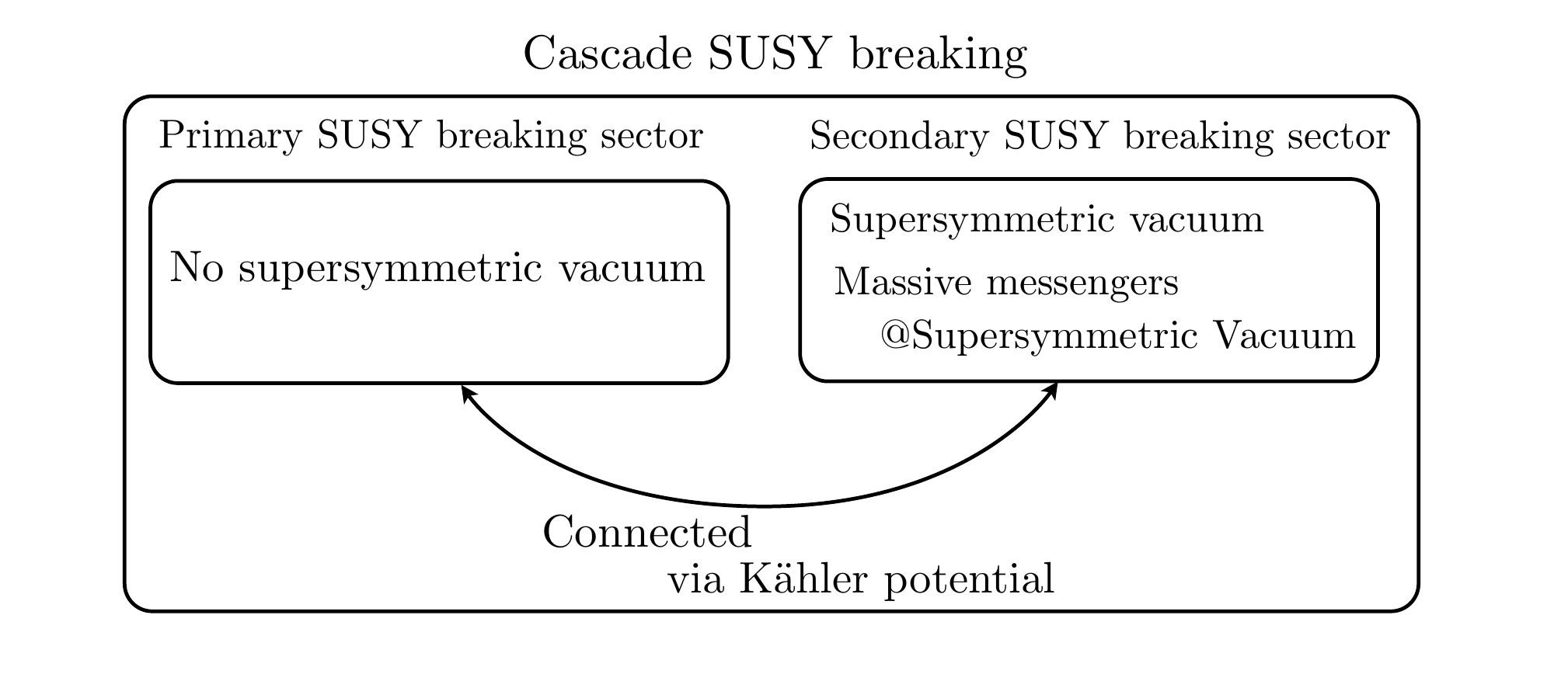}
 \caption{A schematic picture of the cascade supersymmetry breaking.
Supersymmetry breaking in the secondary sector is induced by
the primary supersymmetry breaking via the connections in the K\"ahler potential.
   }
 \label{fig:cascade}
\end{center}
\end{figure}

To illustrate the idea of cascade gauge mediation, we consider a model with the following superpotential
\begin{eqnarray}
\label{eq:simplecascadeW}
 W&=& \Lambda_{1}^2S + \Lambda_{2}^2X-\frac{f}{3}X^3 + k X \bar{\psi} \psi\ .
\end{eqnarray}
Here, $S$ represents a field of the primary SUSY breaking sector, 
while $X$, $\psi$, and $\bar\psi$ are fields in the secondary sector (with latter two serving as messengers). We have assumed that the superpotential couplings between the two sectors are suppressed. In section \ref{sec:model}  we will introduce a dynamical model with absence of the superpotential interactions between the two sectors will be ensured by symmetries. The two sectors will interact only through the \kahler potential\,\footnote{For earlier attempts to connect the dynamical supersymmetry breaking sectors
via the K\"ahler potential, see Ref.\cite{Hotta:1996ag}.} (see Fig. \ref{fig:cascade}):
\begin{eqnarray}
\label{eq:simplecascadeK}
  K&=& |S|^2 + |X|^2 + \frac{c_{SP}}{\L^2} |S|^2 |X|^2 + \cdots\ ,
\end{eqnarray}
In the decoupling limit, $c_{SP}=0$, SUSY is broken in the primary sector, while the 
secondary sector possesses a supersymmetric minimum with spontaneously 
broken R-symmetry\,\footnote{In addition the secondary sector has a supersymmetric charge-color breaking vacuum at $\vev{X}=0$ and 
$\vev{\bar\psi\psi}\neq 0$.
In the next section, we develop a model without supersymmetric charge-color breaking vacua.}:
 \begin{eqnarray}
F_S=\Lambda_1\,,~~~X={\Lambda_2}/{f^{1/2}}\,,~~~\langle \bar\psi\psi\rangle=0\,.
  \end{eqnarray}

Once the coupling between two sectors is turned on, $c_{SP}\neq 0$, supersymmetry breaking in the primary sector
induces supersymmetry breaking in the secondary sector through
higher dimensional operators. Indeed $X$ potential becomes
\begin{eqnarray}
 V = - m_{\rm soft}^2 |X|^2 + |\L_{2}^2 + f X^2 |^2\ ,
\end{eqnarray}
where
\begin{eqnarray}
\label{eq:msoft}
 m_{\rm soft}^2  = c_{SP}\frac{|F_S|^2}{\L_1^2}\ .
\end{eqnarray}
We see that $X$ obtains a non-vanishing $F$-term vev which, in the limit $|m_{\rm soft}|\ll\L_2$, is given by
\begin{eqnarray}
\label{eq:FX}
 F_X = \frac{m_{\rm soft}^2}{2f}\ .
\end{eqnarray}
Now SUSY breaking effects are mediated to the MSSM sector due to the coupling $kX\bar \psi\psi$.
One finds a standard expression for gaugino masses
\begin{eqnarray}
 m_{\tilde g} \sim \frac{\a}{4\pi} \log X|_{\theta^2} \neq 0 \ .
\end{eqnarray}

The very light gravitino is realized  when all the dimensioneless coefficients are of order one while the mass scales in the Lagrangian are comparable, i.e.,
\begin{eqnarray}
\label{eq:limit}
\sqrt F_S \sim \L_{\rm 2}\sim\L_1 \sim m\ . 
\end{eqnarray}
In this limit,
the supersymmetry breaking and the messenger mass scales are of the same order of magnitude, 
\begin{eqnarray}
F_X \sim F_S\ , \quad
\vev{X} \sim \sqrt{F_X}\sim \Lambda_{1,2}\ ,
\end{eqnarray}
as required to obtain light gravitino.

Before closing this section, we comment on the sign of  $c_{SP}$.
The SUSY breaking effects discussed above shift the $X$ vev independently of the sign of $c_{SP}$.
The model may be further simplified when $c_{SP}$ is positive.
In this case, the following superpotential is sufficient to induce SUSY breaking effects in the secondary sector:
\begin{eqnarray}
\label{eq:superP}
 W =\frac{f}{3} X^3 + k X \bar\psi\psi\ .
\end{eqnarray}
There exist a supersymmetric minimum at $\vev X = 0$ where messengers are massless.
However, if $c_{SP}>0$
the supersymmetric vacuum is ``destabilized", and both the scalar and 
the $F$-term components obtain non-vanishing expectation values.

\section{A model of cascade supersymmetry breaking}\label{sec:model} 
The superpotential Eq.(\ref{eq:simplecascadeW}) of our toy model is not the most general one consistent with the symmetry. More importantly, the model possesses a supersymmetric charge-color breaking vacuum. The existence of this vacuum places significant constraints  since it may be hard to ensure sufficiently long lifetime of the SUSY breaking vacuum. Finally,
conditions Eq.(\ref{eq:limit}) must be satisfied in light gravitino scenario. This implies that the model is strongly coupled and requires UV completion.

In this section, we introduce UV complete description of the cascade SUSY breaking
based on the $SP(N_c)\times SU(4)$ gauge theory 
with the matter contents given in Table\,\ref{tab:content}.
The Standard Model gauge groups will be embedded into the global  $SU(5)_{\rm SM}$ symmetry.
We choose tree level superpotential to be
\begin{eqnarray}
\label{eq:totalW}
W =  \l_{ij} S_{ij} Q^i Q^j  + m_R \bar{R} R + m_F \bar{F}F\ ,
\end{eqnarray}
where $\l$'s is a coupling constant, and $m_{R,F}$ denote mass parameters.
Let us first choose parameters that will simplify the analysis of SUSY breaking:
\begin{eqnarray}
\label{eq:validity}
 m_R \gg \Lambda_{1}, \, \Lambda_{2} \gg m_F\ ,
\end{eqnarray}
where $\Lambda_1$ and $\Lambda_2$ are dynamical scales of $SP(N_c)$ and $SU(4)$ groups respectively.
In this regime, we can integrate out massive $\bar R$ and $R$ fields and 
the dynamics of two gauge groups, $SP(N)$ and $SU(4)$, decouples
to the leading order . 
Dynamical scales of the two low energy gauge groups are given by
\begin{eqnarray}
\Lambda^{\prime 2(N_c+1)}_{1}=m_R^4 \Lambda^{2(N_c+1)-4}_{1},~~~\Lambda^{\prime 7}_{2}=m_R^{2N_c}\Lambda_{2}^{7-2 N_c}
\end{eqnarray}
Notice that the $SU(4)$ gauge group is asymptotically free above $m_R$ for $N_c<4$, 
while $SP(N_c)$ group is asymptotically free for $N_c >1$. 

\begin{table}[stdp]
\caption{The field content of the model based on $SP(N_c)\times SU(4)$ 
gauge theory.
Here, we also show global symmetries of the model $SU(5)_{\rm SM}$ and $SU(2(N_c+1))$.
} 
\begin{equation}
 \begin{array}{|c||c|c||c|c| l}
\cline{1-5}
 & SP(N_c)& SU(4)&SU(5)_{\rm SM} & SU(2(N_c+1)) &\\
 \cline{1-5}
\noalign{\vskip-\arrayrulewidth}
\noalign{\vskip-\arrayrulewidth}
\noalign{\vskip-\arrayrulewidth}
\noalign{\vskip-\arrayrulewidth}
 \cline{1-5}
 Q & \mathbf {2 N_c} & \mathbf 1 &\mathbf 1&\mathbf{2(N_c + 1)} &\raisebox{-.7em}[ 0pt ][ 0pt ] {{\Large \} } Primary Sector }\\
S_{ij} & \mathbf 1 & \mathbf 1&\mathbf 1&\mathbf{(N_c + 1)(2N_c+1)}\\
(F, \bar{F}) & (\mathbf 1, \mathbf 1) & (\mathbf {4},  \mathbf {4}^*)  &(\mathbf {5},  \mathbf {5}^*) &
\mathbf{1}&\hspace{.5cm}\text{Secondary Sector}  \\
(R, \bar{R}) & (\mathbf {2N_c}, \mathbf {2N_c})&(\mathbf {4}, \mathbf {4}^*)&\mathbf 1&\mathbf{1}& \hspace{.5cm} \text{Connector} \\
\cline{1-5}
\end{array}%
\nonumber
\end{equation}
\label{tab:content}
\end{table}%

\subsection{Primary supersymmetry breaking sector}
In the limit $m_R\rightarrow\infty$, the $SP(N_c)$ dynamics breaks SUSY through IYIT mechanism\,\cite{Izawa:1996pk} and we will refer to this sector as a primary SUSY breaking sector. Let us briefly review the dynamics of this sector. Below $m_R$ the physics is described by an $SP(N_c)$ gauge group and $N_c+1$ flavors and
a set of gauge singlet fields.
The full superpotential is given by the sum of tree level terms and the quantum constraint:
\begin{eqnarray}
 W=\lambda_{ij}S_{ij}Q_iQ_j +{\cal X}(\mathrm{Pf}(QQ)-{\Lambda^\prime}_{1}^{2(N_c+1)} )\ ,
\end{eqnarray}
where $\cal X$ is a Lagrange multiplier. This superpotential is inconsistent with the supersymmetric ground state%
\footnote{Note that phase transitions are not expected as superpotential parameters are varied \cite{Intriligator:1994sm,Witten:1982df}. Thus, this conclusion is valid for any finite value of $m_R$.}.
A convenient description of the dynamics can be obtained in terms of the meson fields $V_{ij}=Q_iQ_j$. These fields
marry singlets and become massive. Thus, the low energy theory contains only a single gauge singlet field with the superpotential 
\begin{eqnarray}
 W=\lambda \Lambda^{\prime2}_{1} S\ ,
\end{eqnarray}
where $S={\rm Pf}( S_{ij})^{1/(N_c+1)}$.
The \kahler potential has the form
\begin{eqnarray}
 K=|S|^2+\eta_S \frac{|S|^4}{\Lambda^{\prime 2}_{1}}\,.
\end{eqnarray}
For small $\lambda$, the coefficient $\eta_S$ is calculable and negative \cite{Chacko:1998si} ensuring that the vacuum is at the origin:
\begin{eqnarray}
S &=& 0\ ,\cr
F_S & = & \l \Lambda_{1}^{\prime 2}=\l m_R^{4/(N_c+1)}\Lambda_{1}^{(2(N_c+1)-4)/(N_c+1)}\ .
\end{eqnarray}

\subsection{Secondary supersymmetry breaking sector}
Below $m_R$ the secondary sector is described by an s-confining $SU(4)$ gauge 
theory\,\cite{Seiberg:1994bz},
\begin{eqnarray}
\label{eq:secondaryW}
 W = m_F \tr M + \frac{\det M  - \bar B  M B}{\L_2^{\prime 7}}=m_F \tr M + 
\frac{\det M  - \bar B  M B}{m_R^{2N_c}\Lambda_2^{7-2N_c}}\ ,
\end{eqnarray}
where
\begin{eqnarray}
 M_a^{\bar{a}} &=& F_a\bar{F}^a\ . \cr
 B_a  &=& \epsilon_{\alpha_1\cdots \alpha_{4}}  \epsilon_{a a_1\cdots a_{4}}  
 \bar{F}^{\alpha_1 a_1}\cdots  \bar{F}^{\alpha_4 a_4}\ , \cr 
 \bar{B}^{\bar{a}}  &=& \epsilon^{\alpha_1\cdots \alpha_{4}}  \epsilon^{\bar{a} a_1\cdots a_{4}}  
  {F}_{\alpha_1a_1}\cdots  {F}_{\alpha_4a_4}\ ,
\end{eqnarray}
and $\a$'s are the $SU(4)$ indices while $a$ and $b$  are $SU(5)_{\rm SM}$ indices.
We further define rescaled fields, $X$, $\tilde M$, $\tilde B$ and $\tilde{\bar{B}}$
by $X=\sqrt{5/2}\, \mathrm{Tr} M/\Lambda_2^\prime$, $\tilde{M}^a_b=(M^a_b-\mathrm{Tr} M^a_b/5)/\Lambda_2^\prime$,
$\tilde B=B/\L_{2}^{\prime3}$ and $\tilde{ \bar{B}}=\bar B/\L_{2}^{\prime 3}$.

The supersymmetric minimum of the secondary sector is located at
\begin{eqnarray}
\label{eq:susyV}
\vev{X}_{\rm SUSY}  &\simeq&\sqrt{10}  \left(\frac{m_F}{\L^\prime_{2}} \right)^{1/4} \L^\prime_{2} \ , \cr
\tilde{M} &=& 0\ , \cr
\tilde{B} &=& \tilde{\bar{B} }  = 0 \ .
\end{eqnarray}
In this vacuum, all the fields charged under the global $SU(5)_{\rm SM}$ symmetry are massive.%
\footnote{Due to the absence of massless fields charged under $SU(4)$ group, 
the secondary sector does not have charge-color breaking supersymmetric
vacua in the limit $c_{SP}=0$ unlike the toy example of the previous section.
}

For sufficiently small $m_F$, the secondary sector also possesses a metastable SUSY 
breaking vacuum\,\cite{Intriligator:2006dd}.
Typically, models of direct gauge mediation make use of this metastable SUSY breaking vacuum. However, the maximal global symmetry of the secondary sector is spontaneously broken in non-supersymmetric minimum which, in turn, requires a gauge group larger than $SU(4)$ and generically leads to Landau poles for the Standard Model couplings below the GUT scale. We will instead use a supersymmetric vacuum of the secondary sector. The presence of $R$, $\bar R$ fields induces \kahler potential interactions between the primary and secondary sectors and generates non-vanishing $F_X$.

\subsection{Interaction between the two sectors}

\begin{figure}[t]
\begin{center}
  \includegraphics[width=.75\linewidth]{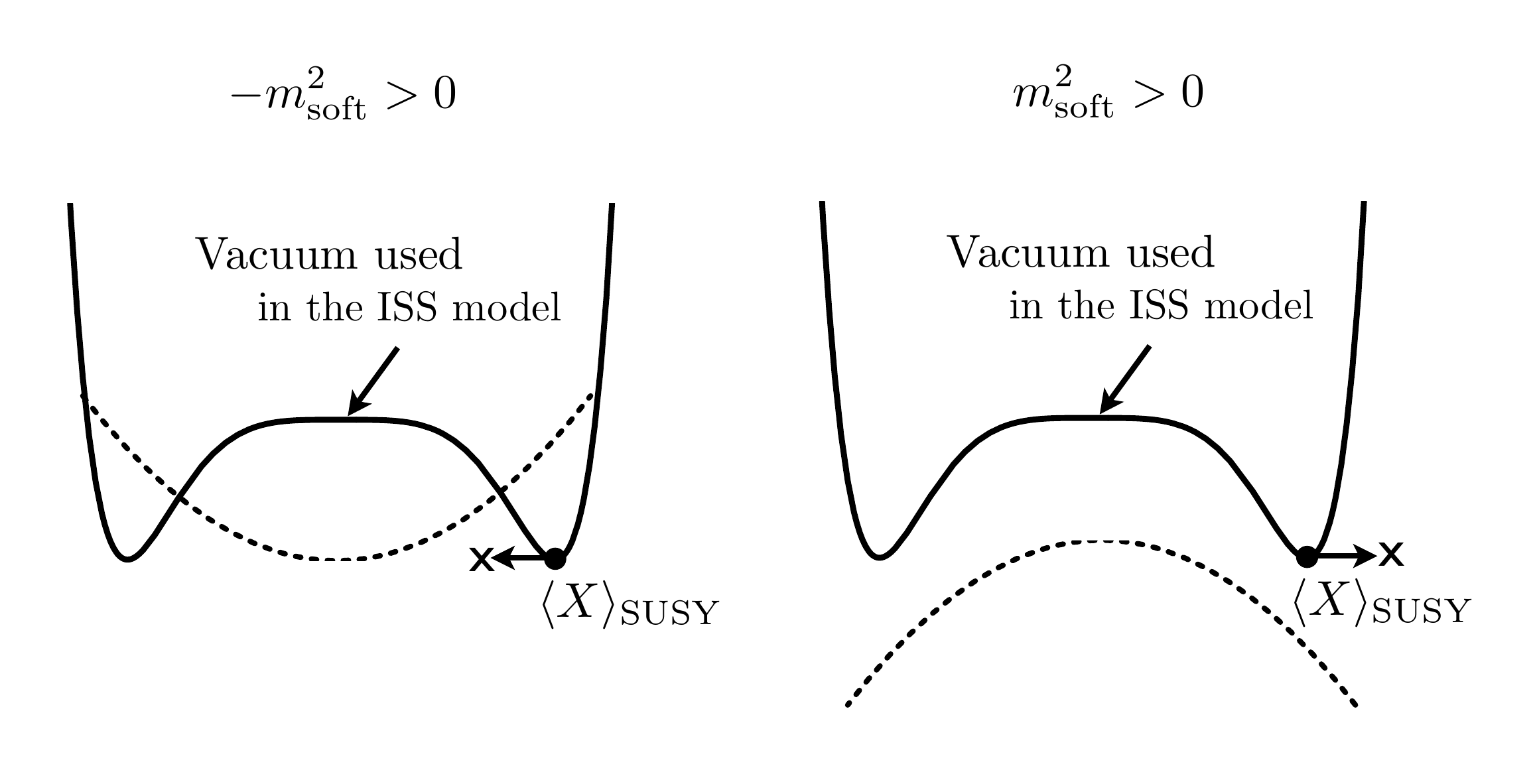}
 \caption{A schematic picture of the shift of the supersymmetric vacuum by the 
 supersymmetry breaking mass via the K\"ahler potential.
 The solid lines show the supersymmetric scalar potential, while  the dashed lines
 show the supersymmetry breaking potential.
 The vacuum used in the cascade supersymmetry breaking is denoted by ``$\times$".
 We have shown a local supersymmetry breaking minimum at $X=0$ discussed 
 in Ref.\cite{Intriligator:2006dd}.
   }
 \label{fig:cascadepot}
\end{center}
\end{figure}

After $R$, $\bar R$ fields are integrated out, the superpotential of the low energy theory is 
\begin{eqnarray}
\label{eq:cascadelowW}
 W & \simeq & \l {\Lambda}_{1}^{\prime 2} S +\frac{5}{\sqrt{10}}\, m_F\L_{2}^\prime X + 
 \frac{1}{10^{5/2}}\frac{X^5}{{\L}_{2}^{\prime 2}}\ ,
\end{eqnarray}
while the interactions between the two sectors are given by corrections to the \kahler potential:
\begin{eqnarray}
\label{eq:cascadelowK1}
 \delta K_1 &\simeq&  c_{SP} \frac{|S|^2 |X|^2 }{m_R^2} + c_{M} \frac{|S|^2 |\tilde M|^2 }{m_R^2}
+ c_{B} \frac{|S|^2 |\tilde B|^2 }{m_R^2}
+ c_{B} \frac{|S|^2 |\tilde{\bar B}|^2 }{m_R^2}\ . 
\end{eqnarray}
It is useful to note that since coupling constants in Eq.(\ref{eq:cascadelowK1}) are radiatively generated, they are small in the large $m_R$ limit. On the other hand, these couplings are of order one in a strongly coupled regime,
 \begin{eqnarray}
\label{eq:strongcoupling}
m_R\sim m_F\sim\Lambda_1\sim\Lambda_2\,.  
 \end{eqnarray}
A comparison of Eqs.(\ref{eq:cascadelowW})  and  (\ref{eq:cascadelowK1}) to Eqs.(\ref{eq:simplecascadeW}) and (\ref{eq:simplecascadeK}) shows that $SP(N_c)\times SU(4)$ model represents a dynamical realization of the cascade supersymmetry breaking.

Let us analyze the spectrum of the model at weak coupling. For $|c_{SP}|\ll 1$, 
the $X$ vev is slightly shifted from its supersymmetric value 
in Eq.(\ref{eq:susyV}),
and the size of the secondary supersymmetry breaking is expected to be much smaller than $F_S$ and 
$\L_{2}^{\prime 2}$.
Furthermore, the effect of the interactions on the vevs of the fields in the primary sector is negligible.
Thus, the effective scalar potential of $X$ can be approximated by%
\footnote{The scalar potential of $X$ possesses
a discrete $Z_4$ symmetry which is a subgroup of the $Z_8$ R-symmetry.
The discrete symmetry is spontaneously broken by $\vev X\neq 0$, which
leads to the domain wall production at the phase transition.
The domain wall is, however, unstable since the $Z_4$ symmetry
is explicitly broken by the constant term in the superpotential through 
the supergravity effects, and hence, it does not cause 
the cosmological domain wall problem\,\cite{Dine:2010eb}.   
}
\begin{eqnarray}
\label{eq:scalarV}
 V \simeq -m_{\rm soft}^2|X|^2 + 
 \left| -\frac{5}{\sqrt{10}} m_F \L^\prime_{2} +\frac{1}{10^{5/2}} \frac{5\,X^4}{\L_{2}^{\prime 2} }\right|^2\ ,
\end{eqnarray}
with 
\begin{eqnarray}
\label{eq:soft}
 m_{\rm soft}^2 ={c_{SP}}\frac{|F_S|^2}{m_R^2} \ .
\end{eqnarray}

As a result, the supersymmetric vacuum in Eq.(\ref{eq:susyV}) is shifted to
\begin{eqnarray}
\label{eq:susyB}
\vev{X }& \simeq&
\vev{X}_{\rm SUSY} \left(1+\frac{1}{4} 
\frac{m_{\rm soft}^2}{\sqrt{m_F^3\L_{2}^\prime }} \right)\ ,\cr
F_X & \simeq &\frac{10^{5/2}m_{\rm soft}^2\L_{2}^{\prime 2}}{20\vev{X}^2}
\simeq \left(\frac{5\L_{2}^\prime}{2m_F}\right)^{1/2} m_{\rm soft}^2\ .
\end{eqnarray}
Here, we have assumed that $|m_{\rm soft}|$ is much smaller than 
$m_F, \L^\prime_{2}$ in the weak coupling regime.

As we have noted before the secondary supersymmetry breaking is achieved regardless of the 
sign of $c_{SP}$. A a schematic cartoon of the supersymmetry breaking 
shift in the scalar potential and its dependence on the sign of $c_{SP}$  
is shown in Fig.\ref{fig:cascadepot}. 
The implications of the possibility of positive $c_{SP}$ are discussed in  the appendix.

\section{Superparnter masses}\label{sec:spectrum}
We are now ready to gauge the Standard Model subgroup of the global $SU(5)_{\rm SM}$ symmetry and consider the effects of supersymmetry breaking on the MSSM sector. We begin by studying the messenger spectrum.
The $SU(4)$ gauge group confines and the composite fields, $\tilde M$ and ($\tilde B$, $\tilde{\bar B}$),
transforming in an adjoint and fundamental representations of the $SU(5)_{\rm SM}$ will serve as messenger fields. 
Messengers obtain both holomorphic and D-type soft masses. The D-type soft masses arise from the \kahler potential interactions with the primary SUSY sector Eq.(\ref{eq:cascadelowK1}) as well as from the \kahler potential terms generated by self-interactions in the secondary sector:
\begin{eqnarray}
\label{eq:cascadelowK2}
 \delta K_2 &=&
\eta_M\frac{|X|^2 |\tilde{M}|^2}{\L_{2}^{\prime 2}} 
+ \eta_B\frac{|X|^2 |\tilde{B}|^2}{\L_{2}^{\prime 2}} 
+ \eta_B\frac{|X|^2 |\tilde{\bar B}|^2}{\L_{2}^{\prime 2}} \ .
\end{eqnarray}
These corrections to the \kahler potential generate D-type scalar masses for messengers:
\begin{eqnarray}
\label{eq:softK}
\left(\tilde{m}_M^2\right)_D &=& - c_{M} \frac{|F_S|^2}{m_R^2} -\eta_M \frac{|F_X|^2} {\L_{2}^{\prime 2}}\ , \cr
\left(\tilde{m}_B^2\right)_D &=& - c_{B} \frac{|F_S|^2}{m_R^2} -\eta_B \frac{|F_X|^2} {\L_{2}^{\prime 2}}\ . 
\end{eqnarray}
Since $c_{M}\sim c_{B}\sim c_{\mathrm SP}$, these terms are of order $m_{\rm soft}^2$.%
Moreover, in the strong coupling regime, self-interactions in the secondary sector result in order 1 corrections to these masses.

In addition messengers receive holomorphic soft masses that can be obtained from 
Eq.(\ref{eq:secondaryW}):
\begin{eqnarray}
\label{eq:softW}
 \left(m_M^2\right)_h &\simeq& \frac{9}{10^{3/2}}\frac{\vev{X}^2 F_X}{\L_{2}^{\prime 2}}\, , \cr
 \left(m_B^2\right)_h &\simeq& \frac{1}{10^{1/2}}F_X\ .
\end{eqnarray}

Finally, supersymmetric terms in the messenger mass matrix are given by
\begin{eqnarray}
m_{M} &\simeq& \frac{3}{10^{3/2}}\frac{\vev{X}_{\rm SUSY}^3}{\L_{2}^{\prime 2}}\ , \cr
 m_{B} &\simeq& \frac{1}{10^{1/2}}\vev{X}_{\rm SUSY}\  . 
\end{eqnarray}
Note that one must choose parameters of the model in such a way that all eigenvalues of the mass squared matrix for scalar messengers are positive. This must be achieved while avoiding too large positive values of D-type soft masses squareds (otherwise gauge mediated masses for the sfermions could become negative).

With the knowledge of the messenger spectrum, we can obtain superpartner masses.
To the leading order in SUSY breaking parameters, gaugino masses only depend on the holomorphic soft masses of the messenger fields\,\cite{Poppitz:1996xw,ArkaniHamed:1998kj} and are given by
\begin{eqnarray}
\label{eq:gaugino}
  m_{\rm gaugino} &\simeq &-\frac{\alpha_a}{4\pi}\left(5\frac{\left(\tilde m_M^2\right)_h}{m_M} + \frac{\left(\tilde m_B^2\right)_h}{m_B}\right)\,.
\end{eqnarray}

A general formula describing soft sfermion masses is presented in Ref.\cite{Poppitz:1996xw}. The contribution due to non-vanishing messenger supertrace dominates the result in the large $m_R$ limit leading to log divergent terms
\begin{eqnarray}
\label{eq:sfermionK}
m_{\rm sfermion}^2 
&\sim& 2 \left(\frac{\alpha_a}{4\pi}\right)^2C_{a}(r)m_{\rm soft}^{2}\left( 
5\log\frac{m_R^2}{m_{M}^2}+\log\frac{m_R^2}{m_{B}^2}\right)\ ,
\end{eqnarray}
where $C_a(r)$ denotes the quadratic Casimir invariant of the MSSM gauge symmetries for each sfermion of a representation $r$ and we made the use of the fact that the D-type masses are of the order $m_\mathrm{soft}$.
Effect of the holomorphic soft messenger masses is small in the large $m_R$ limit,
\begin{eqnarray}
\label{eq:sfermionW}
m_{\rm sfermion}^2 
\sim  2\left(5+\frac{1}{9}\right)\times \left(\frac{\alpha_a}{4\pi}\right)^2C_a(r)\frac{(m_M^2)_h^2}{m_M^2}
\  . 
\end{eqnarray}
To guarantee that the sfermion and gaugino masses are comparable we must take strong coupling\footnote{As is shown in the appendix, this requirement may be somewhat relaxed if $c_{SP}>0$.} limit Eq.(\ref{eq:strongcoupling}). 
In this limit we can only give an estimate of superpartner masses,
\begin{eqnarray}
\label{eq:app1}
 m_{\rm gaugino} &\sim&  \frac{\alpha_a}{4\pi} m_{\rm soft} \ , \cr
  m_{\rm sfermion}^2 &\sim&  \left(\frac{\alpha_a}{4\pi}\right)^2 m_{\rm soft}^2\ .
\end{eqnarray}
Nevertheless this is a desirable region of the parameter space since it results in low scale GMSB.
We also note that 
strong coupling effects could have  significant consequences for  the superpartner spectrum and further suppress the hierarchy between the superpartner masses and fundamental SUSY breaking 
scale\,\cite{Meade:2008wd}. Since such modifications would only improve the plausibility of the light gravitino scenario, we will use the estimate (\ref{eq:app1}) in the rest of the paper.

In table\,\ref{tab:summary}, we show a summary of scales in  model
that allow comparable gaugino and sfermion masses (see the appendix for the detailed analysis).
In the table, we listed the appropriate mass scales for $c_{SP}<0$ and $c_{SP}>0$ separately.
Notice that the sign of $c_{SP}$ is not the parameter of the 
model but is determined by model by model.
Our case study approach just reflects our inability to calculate the sign of $c_{SP}$ due to the strong interactions%
\footnote{It may be possible to modify the sign of $c_{SP}$ by introducing additional interactions in the model.}.

\begin{table}[tdp]
\caption{Summary of the model of cascade supersymmetry breaking based on $SP(N_c)\times SU(4)$ gauge symmetries.
In the parameter regions listed below, the gaugino masses and the sfermion masses 
are comparable.
The discussions for $c_{SP}>0$ are given in the appendix.
The very light gravitino is realized by taking $m_R$ close to $\L_{2}^\prime$ 
in the cases with low scale gauge mediation (see discussion around Eq.(\ref{eq:gravitinomass})).
}
{\small
\begin{center}
\begin{tabular}{|c|c|c|}
 \hline
 & mass parameters & mediation scale \\ 
 \hline
{$c_{SP}<0$}  & 
{$\Lambda_{2}^\prime \sim m_F \sim |m_{\rm soft}|$} 
& 
{low scale} 
\\
 \hline
\raisebox{-.65em}[ 0pt ][ 0pt ]{$c_{SP}>0$}  
& $\Lambda_{2}^\prime \sim |m_{\rm soft}|$ & 
\raisebox{-.65em}[ 0pt ][ 0pt ]{low scale}\\
 &   
 $m_F$ can be small
 & \\
 \hline
 \raisebox{-.65em}[ 0pt ][ 0pt ]{$c_{SP}>0$}  & $\Lambda_{2}^\prime \gg |m_{\rm soft}|$ &
 \raisebox{-.65em}[ 0pt ][ 0pt ]{ high scale}\\
 &   $m_F$ can be small
 & \\
 \hline
\end{tabular}
\end{center}
}
\label{tab:summary}
\end{table}%

\begin{center}
\begin{figure}[t]
\begin{center}
  \includegraphics[width=.8\linewidth]{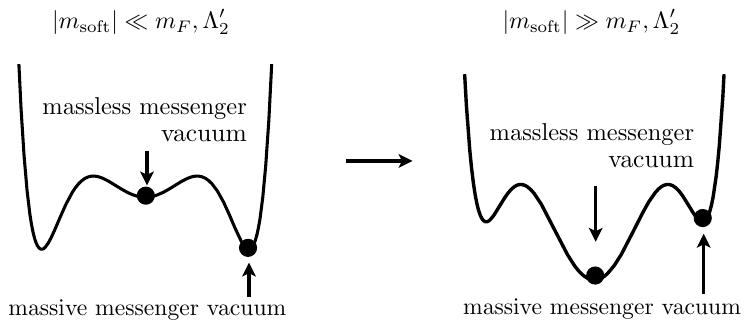}
 \caption{The possible phase transition into the massless messenger  vacuum for $c_{SP}<0$.
 For too large $|m_{\rm soft}|$, the massive messenger vacuum becomes no more
 the global minimum  of the potential.
  }
 \label{fig:phase}
\end{center}
\end{figure}
 \end{center}

As we have explained earlier, the model possesses a local charge-color and SUSY breaking minimum at $X=0$. When $c_{SP}<0$, it is possible that this minimum has lower energy and in the strongly coupled regime the lifetime of the phenomenologically viable vacuum is too short (see Fig.\ref{fig:phase} for a schematic picture).  Let us study this question in the calculable regime.
If $|c_{SP}|\ll 1$, 
the phase transition does not occur as long as the mass parameters  satisfy,
\begin{eqnarray}
 V(0) \simeq \frac{5}{2} 
 m_F^2 \L_{2}^{\prime 2}  \gtrsim |m_{\rm soft}|^2 |\vev{X}|^2 =10 |m_{\rm soft}|^2 (m_F \L_{2}^{\prime 2})^{1/2}\ ,
  \end{eqnarray}
which leads to
 \begin{eqnarray}
 \label{eq:global}
 |m_{\rm soft}|^2 \lesssim \frac{1}{4} (m_F^3\Lambda_{2}^\prime)^{1/2}\ ,
\end{eqnarray}
or
\begin{eqnarray}
\label{eq:upperbound}
\frac{(m_M^2)_h}{m_M} \lesssim \frac{3}{4} |m_{\rm soft}|\ .
\end{eqnarray}
In the final expression, we have used Eqs.(\ref{eq:susyB}) and (\ref{eq:softW}).
Thus, in the calculable regime, we can guarantee the stability of the vacuum
by assuming Eq.(\ref{eq:upperbound}). 
The situation is more complicated in a strongly coupled regime.
In particular, when $m_R\sim\Lambda_2$
composites involving $R$ and $\bar R$ fields can not be integrated out and must be included into effective low energy description. While the vacuum structure may be more complicated in this regime, 
it appears plausible that charge-color preserving vacuum will remain the lowest energy minimum of the potential near the origin of the field space.
In the following, we assume that this is indeed the case.

\subsection{Gravitino Mass}
Let us now obtain a lower bound on the gravitino mass in this model. The fundamental scale of the SUSY breaking is bounded from below by the experimental limits on the sfermion masses, in particular the slepton masses
\cite{Amsler:2008zzb}, 
\begin{eqnarray}
 m_{\tilde \ell} \gtrsim 94\,{\rm GeV}\ .
\end{eqnarray}
In our model we can convert this constraint into the lower bound on soft masses both in the primary and secondary SUSY
sectors:
\begin{eqnarray}
\label{eq:constraint}
 |m_{\rm soft}|\gtrsim 50\,{\rm TeV},~~~~| F_S | \gtrsim c_{SP}^{-1/2} \frac{m_R}{|m_{\rm soft}|} \times \left(50\,{\rm TeV}\right)^2\ ,
\end{eqnarray}
which in turn leads to a lower bound on the gravitino mass\footnote{As mentioned above, the soft masses in Eq.(\ref{eq:app1}) 
could be enhanced by strong coupling effects between messenger fields.
This could allow gravitino to be somewhat lighter.} 
\begin{eqnarray}
\label{eq:gravitinomass}
m_{3/2} = \frac{F_S}{\sqrt{3} M_{\rm PL}} \gtrsim 0.6\,{\rm eV }
\times c_{SP}^{-1/2} \frac{m_R}{|m_{\rm soft}|}\ .
\end{eqnarray}
We see that in the strong coupling limit (\ref{eq:strongcoupling}) the gravitino can be as light as 1 eV.

Before closing this section, let us comment on the effective messenger number of this model.
Messengers are composite fields transforming in the adjoint and 
fundamental representations of the Standard Model guage group. This could lead to the appearance of the Landau pole for the Standard Model gauge coupling below GUT scale \cite{Jones:2008ib}. In our model, however, composite messengers contribute to the renormalization group running of the gauge coupling constants in a interval between the scales $\m_R \simeq m_{M,B}$ and 
$\m_R \simeq \L_{2}^\prime$.
Thus, the effects from the composite messengers are very small for 
$m_F\sim\L_{2}^\prime \sim |m_{\rm soft}| $.
Above the the renormalization scale $\L_{2}^\prime$, on the other hand, 
the model has only four fundamental representations of the $SU(5)_{\rm SM}$ gauge groups, 
and hence, 
the model allows perturbative coupling unification even for low SUSY breaking scale.

\section{Conformal Gauge Mediation}\label{sec:conformal}
The light gravitino mass can be achieved in the model of section\,\ref{sec:model} only if several a priori unrelated mass parameters are comparable.
Here, we show that this apparent coincidence of scales may be a consequence of conformal symmetry of the underlying model (see Ref.\cite{Ibe:2007wp} for the detailed discussion of conformal gauge mediation). Note that in the case where both gauge groups are asymptotically free, $N_c=2,3$, both $SP(N_c)$ and $SU(4)$ are in conformal window. Thus, the existence of the IR conformal fixed point is plausible.
At the fixed point, beta functions for gauge and Yukawa couplings must vanish leading to the following conditions
\begin{eqnarray}
 3 (N_c+1)-(N_c+1)(1-\gamma_Q) -4 (1-\gamma_R) &=&0\ , \cr
  3 \times 4-5(1-\gamma_F) -2 N_c (1-\gamma_R) &=&0\ ,\cr
\gamma_S + 2 \gamma_Q&=&0\ .
\end{eqnarray}
These equations do not uniquely determine anomalous dimensions of the matter fields. 
To do so we use an $a$-maximization method \cite{Intriligator:2003jj}.
The $a$-function of the model is given by,
\begin{eqnarray}
a &=& 2 N_c \cdot 2(N_c+1)\times ( 3 (R_Q-1)^3 -(R_Q-1) )\cr
& &+2 N_c \cdot 2\cdot4 \times( 3 (R_R-1)^3 -(R_R-1) ) \cr
& &+ 10 \cdot 4 ( 3 (R_F-1)^3 -(R_F-1) ) \cr
&&+ (N_c+1)(2N_c +1) ( 3 (R_S-1)^3 -(R_S-1) )\ ,
\end{eqnarray}
where $R_i$'s are related to the anomalous dimensions by $R_i = 2(1+\gamma_i/2)/3$.
The values of anomalous dimensions obtained both by using $a$-maximization method and perturbative one-loop calculation  are presented in the  table\,\ref{tab:cft}. We see a good agreement between the two calculations in the case $N_c=2$, strongly suggesting that a (relatively weakly coupled) fixed point exists. 
In the case $N_c=3$, 
the perturbative calculation breaks down but the existence of the IR fixed point is still plausible.

\begin{table}[tdp]
\caption{Anomalous dimensions at the IR fixed point for $N_c=2,3$. Superscript $p$ denotes results of the perturbative calculation ($N_c=2$)
while the superscript $a$ denotes results obtained by the $a$-maximization method.
For $N_c=3$, the one-loop analysis does not give us meaningful comparison with 
the a-maximization method.}
\begin{center}
\begin{tabular}{|c|c|c|c|c|c|c|c|c|}
\hline
$N_c$ & $\gamma_Q^a$& $\gamma_Q^p$&$\gamma_R^a$& $\gamma_R^p$&$\gamma_F^a$&$\gamma_F^p$& $\gamma_S^a$&$\gamma_S^p$\\
\hline
 $2$ & $-0.078$&$-0.062$ &$-0.44$&$-0.45$ &$-0.25$& $-0.24$&$0.16$&$0.12$\\
 $3$ & $-0.43$&$-$& $-0.57$&$-$& $0.48$& $-$&$0.86$&$-$\\
 \hline
\end{tabular}
\end{center}
\label{tab:cft}
\end{table}%

Let us now assume that, in the UV, the coupling constants are chosen close to the fixed point values while mass parameters $m_R$ and $m_F$ are small but non-vanishing.
The model then quickly flows to the fixed point.
and remains conformal down to scales of order $m_R$ (for $m_R>m_F$). Below $m_R$ conformal symmetry is broken and the low energy physics 
is that of cascade SUSY breaking. If the model is strongly coupled at the fixed point (as is suggested by the values of anomalous dimensions presented in the table \ref{tab:cft}), the scales of low energy $SP(N_c)$ and $SU(4)$ dynamics are expected to be just below $m_R$,
\begin{eqnarray}
 m_R \sim \Lambda_{1}^\prime \sim \Lambda_{2}^\prime.
\end{eqnarray}
Notice that there is no arbitrariness in the relations between the mass scale $m_R$ and the dynamical
scales since the values of the gauge coupling constants at around the energy scale $m_R$
are fixed by the conformal symmetry. 

We still need one coincidence, namely IR values of $m_R$ and $m_F$ must be of the same order. 
It is reasonable to assume that these explicit mass scales are comparable in the UV, $m_R\sim 
m_F\sim m_0$. However, the effects of RG evolution are significant
\begin{eqnarray}
\label{eq:massenhance}
 m_R &=& m_0 \times \left(\frac{m_R}{M_{\rm CFT}} \right)^{\gamma_R}\ ,\cr
 m_F &=& m_0 \times \left(\frac{m_R}{M_{\rm CFT}} \right)^{\gamma_F}\ .
\end{eqnarray}
where $M_{\rm CFT}$ denotes the scale at which the model approaches the fixed point.
Since $\gamma_R<\gamma_F$ we find that $m_F$ is naturally much smaller than $m_R$. 
As discussed in the appendix, the light gravitino scenario is still viable for small $m_F$ if $c_{SP}>0$.

Finally, we note that analysis of the gauge coupling unification becomes more involved in the case of conformal DSB sector.
If the $SU(4)$ gauge coupling is closed to its fixed point value in a large interval of energy scales large anomalous dimensions of the messengers may have a significant effect on the running of MSSM gauge couplings (since the contribution of a single messenger to the beta function is changed by a factor  of $(1-\gamma_F)$).
For $N_c = 3$, the effective number of the messengers is smaller than four due to the 
large positive anomalous dimension of $F$, $\gamma_F = 0.48$.
For $N_c =2$, on the other hand, the number of messengers is enhanced 
and is just below five. 
Thus, careful studies are required to see whether the 
large anomalous dimensions cause the Landau pole problems below the GUT scale, when the scale $M_{\rm CFT}$ is close to the GUT scale.
This problem can be avoided when the scale $M_{\rm CFT}$ is much lower than the GUT scale
and the supersymmetry breaking sector is  weakly coupled above $M_{\rm CFT}$.

Interestingly, when $M_{\rm CFT}$ is small compared to the GUT scale,
it is also possible to relate the MSSM $\mu$-term with the mass parameter $m_0$ required in our model.
If the CFT scale $M_{\rm CFT}=O(10^{10-13})$\,GeV and we choose $\mu\sim m_F\sim m_R\sim m_0\sim (100)\,\mathrm{GeV}-O(1)\,\mathrm{TeV}$, then for $N_c=2$, 
the renormalization group evolution drives $m_R$ (and SUSY breaking scale) to the desired value of  $m_R =O(10-100)$\,TeV while $\mu$ remains at the electroweak scale (see also
\cite{Ibe:2007wp}).

\section{Conclusions and discussion}
In this paper, we proposed a new class of models with gauge mediated supersymmetry
breaking, the cascade supersymmetry breaking, which admits
a low scale gauge mediation and very light gravitino.
This class of models is reminiscent of early GMSB models with several scales in that there is a primary and a secondary SUSY breaking sectors%
\footnote{The latter one was often referred to as a messenger sector}. 
However, in our model supersymmetry breaking 
in the secondary sector is itself achieved through gauge interactions.
As we have demonstrated, it is possible to implement low scale gauge mediation in this class of models while avoiding light gaugino, Landau pole, and vacuum stability problems generic in direct GMSB models.
Furthermore, a specific model presented in this paper may allow gravitino  mass as low as to be as low as
$O(1)$\,eV, in the range that will be probed by the future cosmic microwave background 
observations\,\cite{Ichikawa:2009ir}.

Several comments are in order.
As discussed in Refs.\cite{Hamaguchi:2007rb,Hamaguchi:2008rv, Hamaguchi:2009db,
Yanagida:2010zz},
the models where messengers are charged under the strong gauge dynamics of the DSB sector
generically suffer from the existence of 
the unwanted stable composite fields with MSSM quantum numbers.
The model of section \ref{sec:model} possesses two global $U(1)$ symmetries (under which $F$ and $R$ fields are charged). Unless these symmetries are explicitly broken by higher dimensional terms in the Lagrangian, this implies existence of stable composite particles with masses in the range of tens to hundreds TeV. The annihilation cross-section for these particles may be close to the unitarity bound due to the strong interactions in the DSB sector. Nevertheless, stable composites charged under the Standard Model gauge group are severely constrained both by direct detection experiments and by the Bing Bang nucleosynthesis. 
For example, even the relic density of the neutral component $B$ and $\bar B$
with a mass in the hundreds TeV range
is restricted to be lower than the observed dark matter density by a factor of $O(10^{-4})$\,
\cite{Ahmed:2009zw,Collaboration:2010er}.
Such a low relic density, however, cannot be achieved even when the annihilation
cross section saturates the unitarity limit\,\cite{Griest:1989wd}.

To avoid these constains, at least one global $U(1)$ must be broken explicitly. For example, if $B$ is the lightest composite with the Standard Model quantum numbers, the simplest operator that allows one to avoid direct detection constraints is \,\cite{Yanagida:2010zz},
\begin{eqnarray}
 W = \frac{1}{M_*^2} \bar{F}\bar{F}\bar{F}\bar{F}\bar{5}_{\rm MSSM}\ ,
\end{eqnarray}
where $\bar{5}_{\rm MSSM}$  is the MSSM quarks and leptons in the $\bar{5}$ representation
of $SU(5)_{\rm SM}$, and $M_*$ denotes the ultraviolet cutoff scale.
Here, we have suppressed $SU(4)$ and $SU(5)_{\rm SM}$ indices.
However, if the UV cutoff scale is as high as the GUT or the Planck scale, the lifetime of the composite messengers, is much longer than $1$\,second and
their decay 
may spoil the success of the Big Bang Nucleosynthesis
(see Ref.\cite{Kawasaki:2004qu} for recent constraints).
If, on the other hand, $\bar F R Q$ is the lightest composite with the Standard Model quantum numbers, the $U(1)$ symmetry may be broken by a lower dimension operator
\begin{eqnarray}
  W = \frac{1}{M_*} F\bar{R}Q\bar{5}_{\rm MSSM}\ .
\end{eqnarray}
Since this operator is suppressed by only $M_*$, the resultant lifetime
of the composite particle is much shorter than $1$ second.

A second global $U(1)$ symmetry allows the possibility that a neutral composite, $R^4$ or $\bar R^4$, is also stable. Such a composite provides a plausible dark matter candidate. Despite their large mass the  the relic density of these particles can be consistent with  observations since their annihilation cross section
is expected to be close to the unitarity limit\,\cite{Griest:1989wd}.%
\footnote{
The stable gravitino with mass in the eV range is a sub-dominant component of dark matter.
}

We would also like to note tha in the model of section\,\ref{sec:model} the continuous R-symmetry is broken down to the discrete $Z_8$ R-symmetry by the explicit mass
terms in the secondary sector.
Thus, the model does not have an R-axion even after the $X$ obtains non-vanishing
vev. One exception is the choice of $c_{SP}>0$ which, as discussed in the appendix allows $m_F$ to be small compared to other mass scales in the theory (see table\,\ref{tab:summary}).
In this case, the spontaneous breaking of an approximate continuous R-symmetry leads to
a very light R-axion with mass
\begin{eqnarray}
 m_{a}^2\simeq  \max\left[4 m_F (\L_{2}^\prime m_{\rm soft}^2)^{1/3},\quad
\sqrt{\frac{3}{5}}\frac{m_{3/2}^2 M_{\rm PL}}{(\L_{2}^\prime m_{\rm soft}^2)^{1/3}}  \right]\ .
\end{eqnarray}
Here, the second contribution comes from the explicit R-symmetry breaking 
by the constant term in the superpotential through the supergravity effects\,\cite{Bagger:1994hh}.
Therefore, it is possible that there is a  very light R-axion with 
mass in the hundreds MeV range for $m_F = O(100)$\,keV. Decay of such an axion inside the detector could lead to 
displaced vertex with a low-mass muon pair and  be detactable at the LHC\,\cite{Goh:2008xz}.

Finally, we would like to point out the primary SUSY breaking sector in the model of section\,\ref{sec:model} may give rise to pseudo Nambu-Goldstone bosons with mass in a TeV range. As disscussed in Refs.\,\cite{Ibe:2009dx,Ibe:2009en} these particles may provide another plausible dark matter candidate.
In this case, one can explain the observed excesses of cosmic ray electron/positron fluxes
at  PAMELA\,\cite{Adriani:2008zr},
ATIC\,\cite{Chang:2008zz}, PPB-BETS\,\cite{Torii:2008xu},
and Fermi\,\cite{Abdo:2009zk} experiments, while evading 
the limit on anti-proton flux from PAMELA experiment\,\cite{Adriani:2008zq}.%
\footnote{See also Refs\,\cite{Mardon:2009gw,Mardon:2009rc,Nomura:2008ru} for related studies. }

\section*{Acknowledgements}
We thank Zohar Komargodski for useful comments.
The work of MI and YS was supported by NSF grant by PHY-0653656.
This work of TTY was supported by the World Premier International Research Center Initiative 
(WPI Initiative), MEXT, Japan.
\appendix
\section{Cascade Supersymmetry breaking with $c_{SP}>0$}\label{sec:modelP}
Analysis presented in the main text is valid independently of the sign of $c_{SP}$.
However, in some perturbative models\,\cite{Dine:1993yw,Dine:1994vc,Dine:1995ag,Nomura:1997ur}, 
 $c_{SP}$ is calculable and positive. 
 If this is also the case in a strongly coupled regime of our model,
 a large region of the parameter space leads to viable superpartner spectrum.

This is a consequence of the fact that for positive $c_{SP}$ the phase transition to the vacuum at $X = 0$
does not happen (see Fig.\ref{fig:cascadepot})
even for $(m_F^3\L_{2}^{\prime })^{1/2}<m_{\rm soft}^2$.
The vevs in the ground state are then largely shifted from the supersymmetric vevs in 
Eq.(\ref{eq:susyV}) to
\begin{eqnarray}
\label{eq:positive}
\vev{X} &\simeq& \sqrt{10} (\Lambda_{2}^{\prime2} m_{\rm soft})^{1/3} 
\left(1+ \frac{{1}}{6}
\left(\frac{m_F^3\L_{2}^\prime}{m_{\rm soft}^4}\right)^{1/3} 
+ \cdots
\right)\ , \cr
F_X &\simeq &\frac{\sqrt{10}}{2}(\Lambda_{2}^{\prime2} m_{\rm soft}^4)^{1/3}
\left(
1- \frac{{1}}{3}
\left(\frac{m_F^3\L_{2}^{\prime}}{m_{\rm soft}^4}\right)^{1/3} 
+ \cdots
\right)\ ,\cr
\frac{F_X}{\vev{X}}  &\simeq & \frac{1}{2}m_{\rm soft} \left( 1 + 
 \frac{{1}}{2}
\left(\frac{m_F^3\L_{ 2}^\prime}{m_{\rm soft}^4}\right)^{1/3} 
+ \cdots
\right)\ , 
\end{eqnarray}
and to leading order are independent of $m_F$ which is now a free parameter and, in particular, can be taken small.
One must also verify that the messengers are non-tachyonic for positive $c_{SP}$. A detailed analysis of the messenger mass matrix shows that this is the case when
\begin{eqnarray}
\L_{2}^\prime\gtrsim m_{\rm soft}\ .
\end{eqnarray}

One advantage of the vacuum  Eq.(\ref{eq:positive})
is that the range of allowed values of $m_F$ is much larger 
while the gaugino and sfermion masses can be comparable without additional requirement 
$m_F\sim\L_2\sim m_\mathrm{soft}$. 
In particular, if $c_{SP}$ is positive both low scale GMSB (if $\L_2\sim m_\mathrm{soft}$) 
and high scale GMSB (if $\L_2\gg m_\mathrm{soft})$ can be 
achieved where the gaugino and the sfermions masses are comparable (see table\,\ref{tab:summary})%
\footnote{Of course gravitino can not be very light in high scale GMSB scenario.}.


\begin{thebibliography}{99}  
\bibitem{Dine:1981za}
  M.~Dine, W.~Fischler and M.~Srednicki,
  Nucl.\ Phys.\ B {\bf 189}, 575 (1981);
  S.~Dimopoulos and S.~Raby,
  Nucl.\ Phys.\ B {\bf 192}, 353 (1981);
\bibitem{Dine:1981gu}
  M.~Dine and W.~Fischler,
  Phys.\ Lett.\ B {\bf 110}, 227 (1982);
  Nucl.\ Phys.\ B {\bf 204}, 346 (1982);
  C.~R.~Nappi and B.~A.~Ovrut,
  Phys.\ Lett.\ B {\bf 113}, 175 (1982);
  L.~Alvarez-Gaume, M.~Claudson and M.~B.~Wise,
  Nucl.\ Phys.\ B {\bf 207}, 96 (1982);
\bibitem{Dimopoulos:1982gm}
  S.~Dimopoulos and S.~Raby,
  Nucl.\ Phys.\ B {\bf 219}, 479 (1983).
\bibitem{Affleck:1984xz}
  I.~Affleck, M.~Dine and N.~Seiberg,
  Nucl.\ Phys.\  B {\bf 256}, 557 (1985).
\bibitem{Dine:1993yw}
  M.~Dine and A.~E.~Nelson,
  %
  Phys.\ Rev.\ D {\bf 48}, 1277 (1993)
  [arXiv:hep-ph/9303230].
\bibitem{Dine:1994vc}
  M.~Dine, A.~E.~Nelson and Y.~Shirman,
  %
  Phys.\ Rev.\ D {\bf 51}, 1362 (1995)
  [arXiv:hep-ph/9408384].

\bibitem{Dine:1995ag}
  M.~Dine, A.~E.~Nelson, Y.~Nir and Y.~Shirman,
  %
  Phys.\ Rev.\ D {\bf 53}, 2658 (1996)
  [arXiv:hep-ph/9507378].
\bibitem{Feng:2010ij}
See, for example,  J.~L.~Feng, M.~Kamionkowski, S.~K.~Lee,
  [arXiv:1004.4213 [hep-ph]].
\bibitem{Viel:2005qj}
  M.~Viel, J.~Lesgourgues, M.~G.~Haehnelt {\it et al.},
  Phys.\ Rev.\  {\bf D71}, 063534 (2005).
  [astro-ph/0501562].
\bibitem{Ichikawa:2009ir}
  K.~Ichikawa, M.~Kawasaki, K.~Nakayama {\it et al.},
  JCAP {\bf 0908}, 013 (2009).
  [arXiv:0905.2237 [astro-ph.CO]].

\bibitem{Izawa:1997gs}
  K.~I.~Izawa, Y.~Nomura, K.~Tobe and T.~Yanagida,
  Phys.\ Rev.\  D {\bf 56}, 2886 (1997)
  [arXiv:hep-ph/9705228].

\bibitem{Csaki:2006wi}
  C.~Csaki, Y.~Shirman and J.~Terning,
  JHEP {\bf 0705}, 099 (2007)
  [arXiv:hep-ph/0612241].

\bibitem{Dine:2006xt}
  M.~Dine and J.~Mason,
  Phys.\ Rev.\  D {\bf 77}, 016005 (2008)
  [arXiv:hep-ph/0611312].

  
\bibitem{Shirai:2010rr}
  S.~Shirai, M.~Yamazaki, K.~Yonekura,
  JHEP {\bf 1006}, 056 (2010).
  [arXiv:1003.3155 [hep-ph]].
\bibitem{Kitano:2010fa}
See also a recent review,
  R.~Kitano, H.~Ooguri, Y.~Ookouchi,
  [arXiv:1001.4535 [hep-th]], and references therein.

    \bibitem{Ibe:2005xc}
  M.~Ibe, K.~Tobe, T.~Yanagida,
  Phys.\ Lett.\  {\bf B615}, 120-126 (2005).
  [hep-ph/0503098].
\bibitem{Sato:2009dk}
  R.~Sato and K.~Yonekura,
  JHEP {\bf 1003}, 017 (2010)
  [arXiv:0912.2802 [hep-ph]].
\bibitem{Sato:2010tz}
  R.~Sato, S.~Shirai,
  [arXiv:1005.1255 [hep-ph]].

      

\bibitem{Aaltonen:2009tp}
  T.~Aaltonen {\it et al.}  [CDF Collaboration],
  Phys.\ Rev.\ Lett.\  {\bf 104}, 011801 (2010)
  [arXiv:0910.3606 [hep-ex]].
\bibitem{Hisano:2008sy}
  J.~Hisano, M.~Nagai, S.~Sugiyama {\it et al.},
  Phys.\ Lett.\  {\bf B665}, 237-241 (2008).
  [arXiv:0804.2957 [hep-ph]].
\bibitem{Ibe:2007wp}
  M.~Ibe, Y.~Nakayama, T.~T.~Yanagida,
  Phys.\ Lett.\  {\bf B649}, 292-298 (2007).
  [hep-ph/0703110 [HEP-PH]];
  M.~Ibe, Y.~Nakayama, T.~T.~Yanagida,
  Phys.\ Lett.\  {\bf B668}, 28-31 (2008).
  [arXiv:0802.2753 [hep-th]];
  M.~Ibe, Y.~Nakayama, T.~T.~Yanagida,
  Phys.\ Lett.\  {\bf B671}, 378-382 (2009).
  [arXiv:0804.0636 [hep-ph]].
\bibitem{Shih:2007av}
  D.~Shih,
  JHEP {\bf 0802}, 091 (2008).
  [hep-th/0703196 [HEP-TH]].
\bibitem{Komargodski:2009jf}
  Z.~Komargodski and D.~Shih,
  JHEP {\bf 0904}, 093 (2009)
  [arXiv:0902.0030 [hep-th]].
  
\bibitem{Sato:2009dk}
  R.~Sato and K.~Yonekura,
  JHEP {\bf 1003}, 017 (2010)
  [arXiv:0912.2802 [hep-ph]].
\bibitem{Ibe:2007ab}
  M.~Ibe and R.~Kitano,
  Phys.\ Rev.\  D {\bf 77}, 075003 (2008)
  [arXiv:0711.0416 [hep-ph]].
\bibitem{Giveon:2009yu}
  A.~Giveon, A.~Katz and Z.~Komargodski,
  JHEP {\bf 0907}, 099 (2009)
  [arXiv:0905.3387 [hep-th]].

\bibitem{Giudice:1997ni}
  G.~F.~Giudice and R.~Rattazzi,
  Nucl.\ Phys.\  B {\bf 511}, 25 (1998)
  [arXiv:hep-ph/9706540].


\bibitem{Dine:2007dz}
  M.~Dine and J.~D.~Mason,
  Phys.\ Rev.\  D {\bf 78}, 055013 (2008)
  [arXiv:0712.1355 [hep-ph]].
\bibitem{Carpenter:2008wi}
  L.~M.~Carpenter, M.~Dine, G.~Festuccia {\it et al.},
  Phys.\ Rev.\  {\bf D79}, 035002 (2009).
  [arXiv:0805.2944 [hep-ph]].
\bibitem{Sato:2009yt}
  R.~Sato, T.~T.~Yanagida, K.~Yonekura,
  Phys.\ Rev.\  {\bf D81}, 045003 (2010).
  [arXiv:0910.3790 [hep-ph]].
  

  
\bibitem{Seiberg:2008qj}
  N.~Seiberg, T.~Volansky, B.~Wecht,
  JHEP {\bf 0811}, 004 (2008).
  [arXiv:0809.4437 [hep-ph]].
\bibitem{Randall:1996zi}
  L.~Randall,
  Nucl.\ Phys.\  {\bf B495}, 37-56 (1997).
  [hep-ph/9612426].
\bibitem{Izawa:2005yf}
  K.~-I.~Izawa, T.~Yanagida,
  Prog.\ Theor.\ Phys.\  {\bf 114}, 433-437 (2005).
  [hep-ph/0501254].
\bibitem{Izawa:2008ef}
  M.~Ibe, K.~I.~Izawa and Y.~Nakai,
  Phys.\ Rev.\  D {\bf 80}, 035002 (2009)
  [arXiv:0812.4089 [hep-ph]];
  M.~Ibe, K.~I.~Izawa and Y.~Nakai,
  arXiv:0907.2970 [hep-ph].
\bibitem{Nomura:1997ur}
  Y.~Nomura, K.~Tobe and T.~Yanagida,
  Phys.\ Lett.\  B {\bf 425}, 107 (1998)
  [arXiv:hep-ph/9711220].
\bibitem{Fujii:2003iw}
  M.~Fujii, M.~Ibe and T.~Yanagida,
  Phys.\ Rev.\  D {\bf 69}, 015006 (2004)
  [arXiv:hep-ph/0309064].

\bibitem{Hotta:1996ag}
  T.~Hotta, K.~-I.~Izawa, T.~Yanagida,
  Phys.\ Rev.\  {\bf D55}, 415-418 (1997).
  [hep-ph/9606203]

  
\bibitem{Izawa:1996pk}
  K.~I.~Izawa and T.~Yanagida,
  Prog.\ Theor.\ Phys.\  {\bf 95}, 829 (1996)
  [arXiv:hep-th/9602180];
  K.~A.~Intriligator and S.~D.~Thomas,
  Nucl.\ Phys.\  B {\bf 473}, 121 (1996)
  [arXiv:hep-th/9603158].
  
\bibitem{Intriligator:1994sm}
  K.~A.~Intriligator and N.~Seiberg,
  Nucl.\ Phys.\  B {\bf 431} (1994) 551
  [arXiv:hep-th/9408155].
\bibitem{Witten:1982df}
  E.~Witten,
  Nucl.\ Phys.\  B {\bf 202}, 253 (1982).



\bibitem{Chacko:1998si}
  Z.~Chacko, M.~A.~Luty and E.~Ponton,
  JHEP {\bf 9812}, 016 (1998)
  [arXiv:hep-th/9810253].
\bibitem{Seiberg:1994bz}
  N.~Seiberg,
  Phys.\ Rev.\  {\bf D49}, 6857-6863 (1994).
  [hep-th/9402044].

\bibitem{Intriligator:2006dd}
  K.~A.~Intriligator, N.~Seiberg and D.~Shih,
  JHEP {\bf 0604}, 021 (2006)
  [arXiv:hep-th/0602239].


\bibitem{Dine:2010eb}
  M.~Dine, F.~Takahashi, T.~T.~Yanagida,
    [arXiv:1005.3613 [hep-th]].
\bibitem{Poppitz:1996xw}
  E.~Poppitz and S.~P.~Trivedi,
  Phys.\ Lett.\  B {\bf 401}, 38 (1997)
  [arXiv:hep-ph/9703246].
\bibitem{ArkaniHamed:1998kj}
  N.~Arkani-Hamed, G.~F.~Giudice, M.~A.~Luty {\it et al.},
  Phys.\ Rev.\  {\bf D58}, 115005 (1998).
  [hep-ph/9803290].
\bibitem{Meade:2008wd}
  P.~Meade, N.~Seiberg, D.~Shih,
  Prog.\ Theor.\ Phys.\ Suppl.\  {\bf 177}, 143-158 (2009).
  [arXiv:0801.3278 [hep-ph]].
\bibitem{Amsler:2008zzb}
  C.~Amsler {\it et al.}  [Particle Data Group],
  Phys.\ Lett.\  B {\bf 667}, 1 (2008).
\bibitem{Jones:2008ib}
  J.~L.~Jones,
  Phys.\ Rev.\  D {\bf 79}, 075009 (2009)
  [arXiv:0812.2106 [hep-ph]].

\bibitem{Intriligator:2003jj}
  K.~A.~Intriligator and B.~Wecht,
  Nucl.\ Phys.\  B {\bf 667}, 183 (2003)
  [arXiv:hep-th/0304128].
  
  
\bibitem{Hamaguchi:2007rb}
  K.~Hamaguchi, S.~Shirai and T.~T.~Yanagida,
  Phys.\ Lett.\  B {\bf 654}, 110 (2007)
  [arXiv:0707.2463 [hep-ph]].
\bibitem{Hamaguchi:2008rv}
  K.~Hamaguchi, E.~Nakamura, S.~Shirai and T.~T.~Yanagida,
  Phys.\ Lett.\  B {\bf 674}, 299 (2009)
  [arXiv:0811.0737 [hep-ph]].
\bibitem{Hamaguchi:2009db}
  K.~Hamaguchi, E.~Nakamura, S.~Shirai and T.~T.~Yanagida,
  JHEP {\bf 1004}, 119 (2010)
  [arXiv:0912.1683 [hep-ph]].
\bibitem{Yanagida:2010zz}
  T.~T.~Yanagida, K.~Yonekura,
    [arXiv:1006.2271 [hep-ph]].
    
\bibitem{Ahmed:2009zw}
  Z.~Ahmed {\it et al.}  [The CDMS-II Collaboration],
  arXiv:0912.3592 [astro-ph.CO].

\bibitem{Collaboration:2010er}
  T.~X.~Collaboration,
  arXiv:1005.2615 [astro-ph.CO].

    
\bibitem{Griest:1989wd}
  K.~Griest, M.~Kamionkowski,
  Phys.\ Rev.\ Lett.\  {\bf 64}, 615 (1990).
  
    
\bibitem{Kawasaki:2004qu}
  M.~Kawasaki, K.~Kohri, T.~Moroi,
  Phys.\ Rev.\  {\bf D71}, 083502 (2005).
  [astro-ph/0408426].
\bibitem{Bagger:1994hh}
  J.~Bagger, E.~Poppitz, L.~Randall,
  Nucl.\ Phys.\  {\bf B426}, 3-18 (1994).
  [hep-ph/9405345].
\bibitem{Goh:2008xz}
  H.~-S.~Goh, M.~Ibe,
  JHEP {\bf 0903}, 049 (2009).
  [arXiv:0810.5773 [hep-ph]].
  
\bibitem{Ibe:2009dx}
  M.~Ibe, Y.~Nakayama, H.~Murayama {\it et al.},
  JHEP {\bf 0904}, 087 (2009).
  [arXiv:0902.2914 [hep-ph]].
  
\bibitem{Ibe:2009en}
  M.~Ibe, H.~Murayama, S.~Shirai {\it et al.},
  JHEP {\bf 0911}, 120 (2009).
  [arXiv:0908.3530 [hep-ph]].
\bibitem{Adriani:2008zr}
  O.~Adriani {\it et al.}  [PAMELA Collaboration],
  Nature {\bf 458}, 607 (2009)
  [arXiv:0810.4995 [astro-ph]].
\bibitem{Chang:2008zz}
  J.~Chang {\it et al.},
  Nature {\bf 456}, 362 (2008).
\bibitem{Torii:2008xu}
  S.~Torii {\it et al.},
  arXiv:0809.0760 [astro-ph].
 
\bibitem{Abdo:2009zk}
  A.~A.~Abdo {\it et al.}  [The Fermi LAT Collaboration],
  arXiv:0905.0025 [astro-ph.HE].
\bibitem{Adriani:2008zq}
  O.~Adriani {\it et al.},
  Phys.\ Rev.\ Lett.\  {\bf 102}, 051101 (2009)
  [arXiv:0810.4994 [astro-ph]].


\bibitem{Nomura:2008ru}
  Y.~Nomura, J.~Thaler,
  Phys.\ Rev.\  {\bf D79}, 075008 (2009).
  [arXiv:0810.5397 [hep-ph]].
\bibitem{Mardon:2009rc}
  J.~Mardon, Y.~Nomura, D.~Stolarski {\it et al.},
  JCAP {\bf 0905}, 016 (2009).
  [arXiv:0901.2926 [hep-ph]].
  
\bibitem{Mardon:2009gw}
  J.~Mardon, Y.~Nomura, J.~Thaler,
  Phys.\ Rev.\  {\bf D80}, 035013 (2009).
  [arXiv:0905.3749 [hep-ph]].


  
\end{thebibliography}
\end{document}